\documentclass[12pt]{article}

\usepackage{amssymb,amstext,amsmath,amsthm}
\usepackage[dvips]{graphicx}
\usepackage{latexsym}
\usepackage{psfrag}
\usepackage{amsfonts}
\usepackage{bbm}
\usepackage{floatflt}

\newcommand{\Ref}[1]{(\ref{#1})}

\newcommand{\eqa}{\begin{eqnarray}}
\newcommand{\neqa}{\end{eqnarray}}
\newcommand{\equ}{\begin{equation}}
\newcommand{\nequ}{\end{equation}}
\newcommand{\no}{\nonumber\\}

\def\om{\omega}

\def\la{\langle}
\def\ra{\rangle}
\newcommand{\bra}[1]{\la {#1}|}
\newcommand{\ket}[1]{|{#1}\ra}
\newcommand{\mean}[1]{\la{#1}\ra}

\newcommand{\p}{\partial}

\def\th{\theta}
\def\d{\delta}
\def\f{\frac}
\def\wtl{\widetilde}

\usepackage{bbm}

\newcommand{\SU}{\mathrm{SU}}

\let\eps=\epsilon

\newcommand{\N}{\mathbb{N}}
\def\what{\widehat}

\begin{document}

\title{\Large\bf Physical boundary state for the quantum tetrahedron}

\author{{Etera R. Livine$^a$ and Simone Speziale$^b$\footnote{etera.livine@ens-lyon.fr, sspeziale@perimeterinstitute.ca}} \\
\small{${}^a$\emph{Laboratoire de Physique, ENS Lyon, 
46 All\'ee d'Italie, 69364 Lyon, France}}
\\
\small{${}^b$\emph{Perimeter Institute, 31 Caroline St. N, Waterloo, ON N2L 2Y5, Canada}}}

\date{\small\today}

\maketitle

\begin{abstract}
\noindent We consider stability under evolution as a criterion to select a physical boundary state for the spinfoam
formalism.
As an example, we apply it to the simplest spinfoam defined by a single quantum tetrahedron and
solve the associated eigenvalue problem at leading order in the large spin limit.
We show that this fixes uniquely the free parameters entering the boundary state. Remarkably, the
state obtained this way gives a correlation between edges which runs at leading order with the
inverse distance between the edges, in agreement with the linearized continuum theory. Finally, we
give an argument why this correlator represents the propagation of a pure gauge, consistently with
the absence of physical degrees of freedom in 3d general relativity.
\end{abstract}

\section{Introduction}

LQG is a background independent approach to the quantization of General Relativity (GR)
\cite{book}. In its covariant spinfoam formulation, it provides us with a regularized expression
for the path integral of the full theory. The state of the art in spinfoams is the proposal for
computing the graviton propagator in non-perturbative quantum gravity
\cite{RovelliProp,Io,grav2,3d,grav3,Alesci,4dnum}. One works with a bounded region of space-time and introduces a
semiclassical state peaked on a given boundary geometry. This boundary state
gauge-fixes the spinfoam amplitude \cite{bianca} and allows to compute the
(two-point) correlations for the gravitational field \cite{RovelliProp}, thus
effectively inducing a non-trivial bulk geometry. This can be seen as the proper way to introduce a
background metric in this context of background independent quantum gravity. It further allows a
perturbative expansion (in the inverse of the observation length scale) with a leading order
corresponding to a flat spacetime metric at large scale. The hope is that
this will allow in the long run to perform calculations in quantum gravity and extract useful
phenomenological predictions.

This program has produced a number of interesting results.
First, the graviton propagator point
of view allowed a criticism and a detailed description of the shortcomings of the Barrett-Crane
model \cite{BarrettC}, which has been the most studied spinfoam model up to now \cite{grav2,Alesci,newvertex},
and spurred the developments of promising alternatives \cite{falcao}.
Beside this, a detailed analysis of the perturbative expansion, with both analytical and numerical
computations, was performed in the context of 3d quantum gravity \cite{Io,3d}. Also many
mathematical and technical tools have been developed to compute and study the 4d spinfoam
correlations \cite{grav2,grav3} and there has been a preliminary computation of the three-point
correlation functions \cite{Bianchi}. Finally, there has been a thorough numerical study of the 4d
correlations for the Barrett-Crane model, which confirmed the semiclassical behaviour at large
scale but also allowed to study the short scale behaviour and show that the
correlations get dynamically regularized at the Planck scale \cite{4dnum}.

The boundary state is a key ingredient of this approach. The original ansatz \cite{RovelliProp} is
a Gaussian state with a phase factor in the Hilbert space of boundary spin networks. This ansatz
has been improved in \cite{3d,grav3} to allow an easier study of the perturbative expansion. These
choices do lead to the right behavior for the correlations in the
semiclassical limit, but leave two important questions open:

(i) There are free parameters in the Gaussian ansatz. Can we fix these parameters dynamically?

(ii) The Gaussian state can only be a first order approximation of some physical state in the full
theory. How can we characterize the full boundary state?

We believe that both questions can be addressed taking seriously the fact that we need to work with
a physical state, that is gauge invariant spin network states that solve the quantum gravity
constraints (scalar and momentum constraints). This issue has not been addressed yet in the
literature. We expect that such a dynamical selection of the boundary state will help
characterizing it in the full theory, thus addressing point (ii), but more importantly we expect it
to uniquely fix the parameters of the Gaussian ansatz at leading order, thus addressing point (i)
and increasing the predicting power of the theory.

In this paper, we address these issues in 3d Riemannian quantum gravity. The theory is indeed much simpler in
three spacetime dimensions. Gravity is topological and we know how to spinfoam quantize it exactly
as the Ponzano-Regge model \cite{Ponzano,carloarea}. We can solve this toy model explicitly and we
know the physical states (spin network functionals on the space of flat connections up to
diffeomorphisms). It thus seems to be the perfect arena to address the problem before tackling it
in the framework of 4d spinfoam models.

To keep things simple and explicit, we consider the smallest 3d triangulation, that is a single
tetrahedron. The corresponding Ponzano-Regge spinfoam amplitude is simply Wigner's $\{6j\}$ symbol
for the unitary group $\SU(2)$. Following the setting introduced in \cite{Io,3d}, we consider the
``time-fixed" tetrahedron: out of its six edges, we study the correlations between the fluctuations
of two opposite edges while freezing the lengths of the remaining four edges. The fixed-length
edges define the time interval associated to that piece of 3d spacetime, while the two fluctuating
edges are distinguished as the initial and final edges. Carefully defining a physical state as a
wave function unaffected by time evolution, we show that the phased Gaussian ansatz actually turns
out to be a true physical state at first order in the asymptotical regime.

More precisely, the physical state criterion uniquely fixes the width of the Gaussian. First it
fixes the (leading order) scaling of the Gaussian width with the observation length scale, which
turns out to be the right scaling in order to recover a flat metric in the semi-classical regime
\cite{Io,3d,semi}. Second it fixes the exact factor in front of the scaling, which turns out to be such
that the leading order of the graviton propagator goes precisely as the inverse distance between the edges.

Finally, we look at the tetrahedron from a covariant point of view. There is a unique physical
state, namely the flat connection boundary state. We discuss the gauge-fixing procedure to go down
to the ``time-fixed" tetrahedron. We also address the issue that there is actually no
graviton in 3d gravity (since it is a topological theory without local degree of freedom) and
argue that the  3d spinfoam graviton correlations entirely come from the choice of gauge
fixing.

\section{From the propagation kernel to the boundary state}
To better understand the physics behind this paper, let us briefly recall
some basic facts of quantum mechanics and quantum field theory, concerning the
propagation kernel and the vacuum state, in addition to their physical interpretation.
These objects incorporate key aspects of the interpretative problems of
background independent quantum gravity.

Consider a scalar $\phi$ in the Schr\"odinger picture \cite{Symanzik, Rossi, Mattei}, and introduce
two spacelike hyperplanes in Minkowski spacetime, separed by a time $T$. Denote $\phi_1$ and $\phi_2$
two classical field configurations associated with these planes,
and $\psi_n[\phi]$ a complete  basis of energy eigenstates.
Given a state $\psi[\phi_1] = \sum_{n} c_n \psi_n [\phi_1]$ on the initial plane,
the evolution to the final state can be written
in terms of the propagation kernel $K[\phi_1, \phi_2, T]$ through
\equ\label{evo1}
\psi(\phi_2, T) = \int {\cal D}\phi_1 \ K[\phi_1, \phi_2, T] \ \psi(\phi_1, 0)
 = \sum_{n} c_n \, e^{-i E_n T} \, \psi_n [\phi_2,0].
\nequ
The second equality shows that the kernel (that can be evaluated as a path integral with $\phi_1$ and $\phi_2$ as boundary data) can be decomposed as
\equ\label{K}
K[\phi_1, \phi_2, T] = \sum_{n} e^{-i E_n T} \psi_n [\phi_1] \overline{\psi_n [\phi_2]}.
\nequ
Among the $n$-particle states in \Ref{K}, the vacuum $\psi_0[\phi]$ is characterized by having
the minimal energy. It can be extracted directly from the kernel using a Wick rotation to
Euclidean time $T_{\rm E} = i T$,
\equ\label{proj}
\lim_{T_{\rm E}\mapsto\infty} K[\phi,0,-iT_{\rm E}] = \psi_0[\phi].
\nequ

Finally, using the kernel \Ref{K} and vacuum \Ref{proj} one can formally rewrite the 2-point function as
\eqa\label{2point}
W(x_1, x_2) &=& \f1{\cal N}\int {\cal D}\phi \, \phi(0,\vec x_1)\, \phi(T,\vec x_2)\,e^{-S[\phi]} = \no &=&
\f1{\cal N} \int {\cal D}\phi_1 \, {\cal D}\phi_2 \, \phi_1(\vec x_1)\, \phi_2(\vec x_2)\,
\psi_0[\phi_1] \, \psi_0[\phi_2] \, K[\phi_1,\phi_2,T].
\neqa
For the sake of the following discussion, let us stress that this gives the 2-point function in vacuum,
$\bra{0} \phi(x_1) \phi(x_2) \ket{0}$. The same formalism can be applied to more general cases,
such as for instance the 2-point function at finite temperature $\tau$, in which case instead of $\psi_0[\phi]$
in \Ref{2point} one should use the appropriate thermal state $\psi_\tau[\phi]$.
A case of interest in the following is when the state is a coherent state of the theory.

These are all well-known results from QFT that crucially use the presence of the Minkowski
background. In particular, they can be extended to linearized GR around Minkowski, using the temporal gauge
\cite{Mattei}.
On the other hand, the situation is more subtle in full GR, where all the physical states have to solve
the Hamiltonian constraint and thus have zero energy. Therefore
the kernel can not have a decomposition like \Ref{K}, but rather of the following type,
\equ\label{KGR}
K[g_1, g_2] = \sum_{n} \psi_n [g_1] \overline{\psi_n [g_2]}.
\nequ
Here $n$ labels a complete basis of physical eigenstates, and it could be a continuous label
(in which case the summation above is truly an integration). In this expression, the physical time
is contained in the metric.\footnote{Assuming that \Ref{KGR} is peaked on a classical trajectory
$g_{\mu\nu}$ between $g_1$ and $g_2$, the physical time can be computed evaluating the $g_{00}$ component
along a geodesic, see discussion in \cite{Mattei}.}
Consequently, the evolution equation \Ref{evo1} becomes a stability condition that any physical state has to satisfy,
\equ\label{evo}
\psi_{\rm ph}[g_2] = \int {\cal D}g_1 \ K[g_1, g_2] \ \psi_{\rm ph} [g_1].
\nequ

Equations like \Ref{KGR} or \Ref{evo} encode the core of GR, and a successful theory of quantum gravity
should be able to realize them explicitly.
The spinfoam formalism provides us with a definition of
the non-perturbative path integral which can be used to investigate \Ref{KGR} and \Ref{evo}.
Here we provide a simple toy model in which we can realize them exactly.
Furthermore, this explicit model will allow us to address another key issue, which concerns the interpretation
of \Ref{2point} in quantum gravity. As thouroughly discussed in \cite{grav2}, in full GR there is no
special vacuum state singled out from the kernel, thus we can not apply the definition
\Ref{2point} straighforwardly. It is more sensible to consider the 2-point function on a semiclassical
state $\Psi_q[g]$, defined as a physical state peaked around a classical metric $q$.
Then one can use a formula like \Ref{2point}, where instead of the ambiguous vacuum state
$\psi_0[g]$, one uses a semiclassical state,
\equ\label{Wq}
W_q(x_1, x_2) = \f1{\cal N}
\int {\cal D}g_1 \, {\cal D}g_2 \, g_1(\vec x_1)\, g_2(\vec x_2)\,
\Psi_q[g_1] \, \Psi_q[g_2] \, K[g_1,g_2].
\nequ
Upon suitable gauge-fixing,\footnote{The issue of gauge-fixing is explicitly under control in the
linearized theory \cite{Mattei}. There one evaluates the kernel and boundary states in the
temporal gauge. Then the presence in \Ref{Wq} of the field insertions requires an additional time-independent gauge-fixing, such as the Coulomb gauge.} this is a well-defined prescription, and indeed one can consider the
2-point function on a semiclassical state also in conventional QFT.
However, this prescription requires the explicit knowledge of $\Psi_q$.
To be a physical semiclassical state, it has to (i) satisfy \Ref{evo}, and (ii)
be peaked around a classical geometry $q$ with minimal uncertainty.
As discussed in \cite{RovelliProp}, such a state should be peaked around both the intrinsic
${R}_{ij}(q)$ and extrinsic ${\cal K}_{ij}(q)$ curvatures of $q$, so formally
\equ\label{RK}
\mean{\widehat R_{ij}} \equiv \int {\cal D}g \, \overline{\Psi_q[g]} \, \widehat R_{ij} \, \Psi_q[g] = R_{ij}(q),
\quad
\mean{\widehat {\cal K}_{ij}} \equiv \int {\cal D}g \, \overline{\Psi_q[g]} \, \widehat {\cal K}_{ij}\,\Psi_q[g]={\cal K}_{ij}(q).
\nequ
The latter are in general complicate equations to define and attempt to solve. To by-pass this difficulty,
an interesting alternative is to find the analog of \Ref{proj} in background independent
quantum gravity, so to be able to extract the boundary state directly from the kernel.

In the rest of this paper, we investigate these ideas in the simple toy model provided by a 3d Riemannian
quantum tetrahedron. Using the spinfoam formalism, we give an explicit realization of \Ref{KGR} or \Ref{evo},
which we use to study the relation between the kernel and the boundary state, as well as its consequences on \Ref{Wq}.

\section{The tetrahedron}
\begin{floatingfigure}[r]{6cm}
\includegraphics[width=4cm]{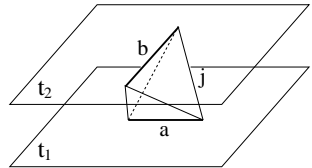}
\caption{\footnotesize The tetrahedron.\label{totti}}
\end{floatingfigure}
\noindent We consider the quantum tetrahedron introduced in \cite{Colosi,Io}, see Fig.\ref{totti}.
Given a tetrahedron of edge lengths $\ell_e$, we fix four opposite ones to $j$, and the remaining two to $a$ and $b$.
We orient it as in Fig.\ref{totti} in such a way that we can think of it as representing the evolution
of the edge $a$ into the edge $b$, in a ``time'' $j$.

The dynamics of this model was studied at both the classical and quantum level in \cite{Colosi}.
The classical dynamics is encoded in the Regge action $S_{\rm R}(a,b,j) = \sum_e \ell_e \theta_e(\ell)$, where
$\theta_e$ are the dihedral angles of the tetrahedron.
The quantum dynamics can be studied \`a la Ponzano-Regge \cite{Ponzano}, making the ansatz that the
lengths in the quantum theory can only have half-integer values $\ell_e = j_e+\f12$ with $j_e \in
{\mathbbm N}/2$, and associating with the tetrahedron the amplitude
\equ\label{Ktet}
K[a, b, j] = \sqrt{d_a \, d_b} \left\{ \begin{array}{ccc} a & j & j \\ b & j & j \end{array}\right\}.
\nequ
In this expression $d_a\equiv 2a+1$ is the dimension of the spin-$a$ representation of $\SU(2)$ and
the $\{6j\}$ is Wigner's 6j-symbol for the recoupling theory of SU(2). The simple form \Ref{Ktet}
allows us to study explicitly if a formula like
\Ref{KGR} is indeed realized, and what its connection with semiclassical states might be. In
particular, the stability condition for physical states \Ref{evo} simply reads
\equ\label{evolution}
\psi(b) = \sum_{a} K[a, b, j] \, \psi(a).
\nequ

Before proceeding, notice that if we view this tetrahedron as part of
a triangulation of flat 3d space between the two planes, we can also introduce the ``asymptotic time''
$$
T(j)=t_2-t_1 = \f{6 V(a,b,j)}{ab},
$$
where
\equ\label{V}
V(a,b,j) = \f1{12}\sqrt{4 a^2 b^2 j^2 - a^2 b^4 - a^4 b^2}
\nequ
is the volume of the tetrahedron. 
In the isosceles case $a=b=j_0$, we have $T=\sqrt{(2j^2-j_0^2)/2}$ and $V=j_0^2 T /6$.
The latter quantities will frequently enter our calcuations below.
Notice that fixing the bulk edges (to $j$) we did not fix the asymptotic time $T$,
but we indeed fixed the way $T$ depends on the (intrinsic and extrinsic) boundary geometry.
This fixing can then be viewed as the analog of the temporal gauge fixing used to define \Ref{KGR} in full GR,
see \cite{Mattei}.

\vspace{0.5cm}

\subsection{Diagonalizing the kernel}
Consider the kernel as a $d_j$-by-$d_j$ matrix $K_{ab}[j] = K[a, b, j]$. Unexpectedly,
this matrix satisfies $K^2\equiv \mathbbm 1$ for all $j$, thus the evolution generated by it is
unitary.\footnote{This might generate some confusion, as it is often stated that evolution in quantum gravity can not
be unitary, due to the absence of a physical clock variable. However, the model here considered has a preferred
clock variable built in it, and the evolution is unitary with respect to it.}
As $K^2\equiv \mathbbm 1$, $K$ is diagonalizable and all its eigenvalues are $\eps_n=\pm 1$.
Using the notation $\psi_{n}(a)=\bra{a}n\ra$ to indicate the $n$-th eigenvector in the basis $a$, we have
\equ\label{decomp}
K_{ab}[j] = \sum_{n=0}^{2j} \eps_n \psi_{n}(a) \overline{\psi_{n}(b)}.
\nequ
This is an intriguing formula: in spite of the simplicity of the model, gives a non-trivial and explicit realization
of \Ref{KGR} in quantum gravity. With respect to \Ref{KGR}, notice the presence here of two possible eigenvalues
$\pm1$. This is the usual manifestation of the fact that the PR model sums over both orientation of the tetrahedron.

The next step is to work out the eigenvectors solving the associate eigenvalue problem,
\equ\label{eigen}
\sum_{b=0}^{2j} K_{ab}[j]  \psi_{j}(b) = \pm \psi_j(a)
\nequ
It would be important to find the general solution to this problem,
but we have only partially succeeded in doing so.
It is easy to show that for any value of $j$,
$\psi_0(a) = \sqrt{d_a}$ is an eigenvector with eigenvalue $\eps_0=1$.
This can be done using the integral representation of the $\{6j\}$,
\begin{equation}\label{integral}
\left\{
\begin{array}{ccc} a & j & j \\ b & j & j \end{array} \right\} =
\int dg_1 \, dg_2\ \chi_{a}(g_1) \, \chi_{b}(g_2) \, \chi_{j}(g_1g_2) \, \chi_{j}(g_1g_{2}^{-1}).
\end{equation}
Then, using the fact that $\sum_b {d_b} \, \chi_{b}(g) = \d(g)$, one can immediately show that
$\sum_{b} W_{ab} \psi_0(b) = \psi_0(a)$.
Similarly one can show that $\psi_1(a) = (-1)^{a}/{\sqrt{d_a}}$ is an eigenvector with
eigenvalue $\eps_1=~(-1)^{2j}$.
The remaining eigenvectors have to be found on a case by case basis for fixed $j$.
For instance for the case $j=1$ we have
$$
\psi_0(a) = (1,\sqrt{3}, \sqrt{5}), \qquad \psi_1(a) = (1,-\f1{\sqrt{3}}, \f1{\sqrt{5}})
$$ with eigenvalue 1, and $\psi_2(a)=\eps_{abc} \psi_0(b) \psi_1(c)$ with eigenvalue $-1$.

For the general case, we can rewrite the eigenvalue problem \Ref{eigen} as an integral equation
using the expression \Ref{integral} as shown in Appendix A. However at this stage we do not have
explicit solutions, so we can not study exact physical semiclassical states as linear combinations
of eigenstates. In the next section we show how to solve it approximately.

\subsection{Semiclassical states}
The perturbative expansion we consider is the large spin limit (in which the half-integers approximate real lengths),
where the $\{6j\}$ symbol is dominated by exponentials of the Regge action,
\equ\label{asymp}
\{6j\} \sim \f1{\sqrt{12\pi V(a,b,j)}}\,\cos\left( S_{\rm R}[a,b,j]+\f\pi4 \right).
\nequ
This is indeed the property that makes it possible to show that the quantum theory based upon
the $\{6j\}$ has a sensible semiclassical limit \cite{Ponzano, altri, Io, 3d, miei}.

The lowest order of $\Psi_q$ should satisfy \Ref{eigen} with the linearized kernel and be peaked around $q$.
In the simple setting considered here, the intrinsic and extrinsic geometry of $q$ are specified
giving a value $j_0$ for edge length and a value $\theta_0$ for its dihedral angle.
We choose the latter in such a way that the complete background tetrahedron is given by the isosceles
configuration $a=b=j_0$. For the exterior dihedral angles associated to $j_0$ and $j$,
elementary geometry gives respectively
\equ\label{coseni}
\cos\theta_0 = -\f{4j^2-3j_0^2}{4j^2-j_0^2},
\qquad
\cos\theta = -\f{j_0^2}{4j^2-j_0^2}.
\nequ
By analogy with the continuum, we expect \Ref{RK} to be implemented in this approximation by a Gaussian
state around $q=(j_0,\theta_0)$, for which we make the following ansatz,
\equ\label{ansatz}
\Psi_{q}(a) = N \, \exp\{-\f1{4\sigma}(a-j_0)^2\} \, \cos\Big(\f{d_a}2 \theta_0 + \phi \Big).
\nequ
Here $N$ is the normalization, $\sigma \in {\mathbbm R}_+$ the width
and $\phi$ is a phase that we leave undetermined for the moment. For $\sigma$ scaling linearly
with the spins, this Gaussian is peaked on $q$ in the large spin limit (e.g. \cite{semi}).
This kind of states have been extensively used as ansatz for leading order semiclassical states
in the recent spinfoam literature \cite{RovelliProp,Io,grav2,3d,grav3,Alesci,Bianchi,semi,Magliaro}.
Yet, a semiclassical state typically has a precise width uniquely fixed by the dynamics
(e.g. the factor $\om$ in the exponent of the vacuum Gaussian of the harmonic oscillator).
Thus a crucial question is whether also in spinfoams the width can be fixed requiring the
state to solve the dynamics. Here we address the question in the specific model \Ref{eigen}, and find a positive answer:
as we show below, \Ref{ansatz} is an eigenstate of the kernel for a unique choice of the width, given by
\equ\label{sigma}
\sigma = \f{4j^2-j_0^2}{8 j} = - \f{j_0^2}{8j} \cos\theta  > 0,
\nequ
and for two choices of $\phi$, corresponding to eigenvalues $\pm1$:
\eqa\label{plus}
\sum_a K[a,b,T] \, \Psi_q(a) &\simeq& \Psi_q(b) \qquad {\rm for} \qquad \phi = d_j \theta + \f14(\pi-\theta), \\
\sum_a K[a,b,T] \, \Psi_q(a) &\simeq& -\Psi_q(b) \qquad {\rm for} \qquad \phi = d_j \theta + \f14(\pi-\theta) +\pi.
\label{minus}\neqa
Here $\simeq$ means at first order in the large spin limit.
The presence of both eigenvalues could be anticipated from \Ref{decomp}, and show that both approximate
eigensolutions can be constructed as linear superposition of exact eigenvectors restricted to the same 
eigenvalue.

Proving these results is straighforward, as  in the large spin limit
the eigenvalue problem \Ref{eigen}, using \Ref{asymp} and
the ansatz \Ref{ansatz}, amounts to a simple evaluation of Gaussian integrals.
However, there is a clear restriction that should be kept in mind: formula \Ref{asymp}
holds only if the volume $V$ appearing in it is real, namely if $V^2$ defined in \Ref{V} is positive.
For a generic configuration $(a,b,j)$ this is not always the case. The quantum range of $a$ and $b$
is $[0, 2j]$, and the condition $V^2>0$ is violated when the endpoints are approached by one of the variables.
Specifically, we have  $V^2 \sim 0$ for $a$ and/or $b$ close to zero, and $V^2 < 0$ for $a$ and/or
$b$ close to $2j$.
In both cases the asymptotics is not \Ref{asymp} (we report the relevant formulas in Appendix B for completeness).
The calculations show below thus only apply in the regime
$a \sim b \sim j$ where we can use \Ref{asymp}.\footnote{A more thorough
approach could also be possible: in \cite{Schulten} a unique expression that interpolates between
the three regimes ($V^2>0, \sim 0, <0$) was given. However it leads to a more complicate analysis,
which we do not attempt here.}

We now follow a standard procedure (e.g. \cite{semi}), but with particular care about the phase.
In the large spin limit, we can approximate the summation
in \Ref{eigen} with an integral, and using \Ref{asymp} we have
\eqa\label{1}
\sum_a K[a,b,T] \, \Psi_q(a) &\simeq& \f{N}4 \sum_{\eps,\eta=\pm} \int da\ \sqrt{\f{d_a \, d_b}{12 \pi V(a,b,j)}} \times
\\\nonumber &&
\exp\Bigg\{i \eps \Big(S_{\rm R}(a,b,j)+\f\pi4\Big) - \f1{4\sigma}(a-j_0)^2 +i \eta\Big(\f{d_a}2\theta_0 + \phi \Big)
\Bigg\}.
\neqa
Next, we expand the Regge action around the background $q$ to quadratic order, obtaining
\equ\label{Rexp}
S_{R}(a,b,T) = 2 d_j \theta + \f{d_a}2 \theta_0 + \f{d_b}2 \theta_0 + \f12 \, G_{lm} \, \d j_l \, \d j_m,
\nequ
where we introduced the shorthand notation $\d j_l = (a-j_0, b-j_0)$. The second derivatives
of the Regge action where computed for instance in \cite{3d}, and evaluated on $q$
give $G_{11} = G_{22} = 
{\cos\theta}/{T}$
and $G_{12} = G_{21} = 
-1/{T}$, where we recall $T=\sqrt{(2j^2-j_0^2)/2}$.

The first three terms of \Ref{Rexp} contribute to the overall phase of \Ref{1}, which reads
\equ
i \Big[(\eps + \eta) \f{d_a}2 \theta_0 + \eps \Big(\f{d_b}2 \theta_0 + 2 d_j \theta + \f\pi4\Big) + \eta \phi\Big].
\nequ
The presence of a phase term in $a$ exponentially suppresses the summation in \Ref{1}.
Therefore the configurations $\eta = -\eps$ for which the $a$-phase vanishes dominate \Ref{1},
and we can write
\eqa\label{2}
\sum_a K[a,b,T] \, \Psi_q(a) &\simeq& \f{N}4 \sum_{\eps = \pm} \int da\ \sqrt{\f{d_a \, d_b}{12 \pi V(a,b,j)}} \times
\\\nonumber &&
\exp\Bigg\{i \eps \Big[\f{d_b}2 \theta_0 + 2 d_j \theta + \f\pi4 - \phi\Big] - \f1{4\sigma}(a-j_0)^2
+i \f\eps2 \, G_{lm} \, \d j_l \, \d j_m \Bigg\}.
\neqa
This is a Gaussian integral whose leading order is obtained at
\equ\label{m0}
\sqrt{\f{d_a \, d_b}{12 \pi V(a,b,j)}}\Big|_{a=b=j_0} = 
\sqrt{\f{2}{\pi T}}.
\nequ
Then
\equ\nonumber
\int da\ \exp\Bigg\{ -\f1{4\sigma}(a-j_0)^2 +i \f\eps2 \, G_{lm} \, \d j_l \, \d j_m \Bigg\} =
2 \sqrt{\f{\sigma \pi}{1-2 i \eps \sigma G_{11}}} \exp\Bigg\{ -\f1{4\sigma'}(b-j_0)^2 \Bigg\}
\nequ
where $\sigma'$ a complex function of $\sigma$, $\eps$, $G_{11}$ and $G_{12}$ that can be easily computed.
In particular, requiring $\sigma'$ to be real implies
$4\sigma^2 = 4\sigma'{}^2 = (G_{12}^2-G_{11}^2)^{-1} = T^2/\sin^2\theta$,
independently of $\eps$. Plugging the values for the second derivatives of the Regge action
given above we obtain the result \Ref{sigma}.
Using this result, a little algebra gives
\equ
\sqrt{\f{\sigma \pi}{1-2 i \eps \sigma G_{11}}} = \sqrt{\f{\pi T}2} 
\exp\Big\{ i {\eps}\Big(\f\pi4 - \f{\theta}2 \Big) \Big\}.
\nequ
The modulus of this quantity exactly cancels the contribution \Ref{m0}, so \Ref{2} reads
\equ\label{3}
\sum_a K[a,b,T] \, \Psi_q(a) \simeq \f{N}{2}
\sum_{\eps \pm} \exp\Bigg\{i \eps \Big[\f{d_b}2 \theta_0 + 2 d_j \theta + \f{\pi-\theta}2 - \phi \Big] - \f1{4\sigma}(b-j_0)^2 \Bigg\}.
\nequ
Thanks to the independence of $\sigma$ from $\eps$ the summation in \Ref{3} gives a cosine and we finally get
\equ\label{4}
\sum_a K[a,b,T] \, \Psi_q(a) \simeq N \,
\exp\{ - \f1{4\sigma}(b-j_0)^2 \} \, \cos\Big(\f{d_b}2 \theta_0 + 2 d_j \theta + \f{\pi-\theta}2 - \phi \Big).
\nequ
At this point it is easy to fix $\phi$ to obtain a result proportional to \Ref{ansatz}:
the choice $\phi = d_j \theta + (\pi-\theta)/4$ leads to an eigenstate with eigenvalue $+1$,
whereas the choice $\phi = d_j \theta + (\pi-\theta)/4 +\pi$ leads to an eigenstate with eigenvalue $-1$.
With these choices of $\phi$ and with $\sigma$ given in \Ref{sigma}, \Ref{ansatz} are approximate eigensolutions
of \Ref{eigen} in the large spin limit.

\begin{figure}[ht]\begin{center}
\includegraphics[width=7cm]{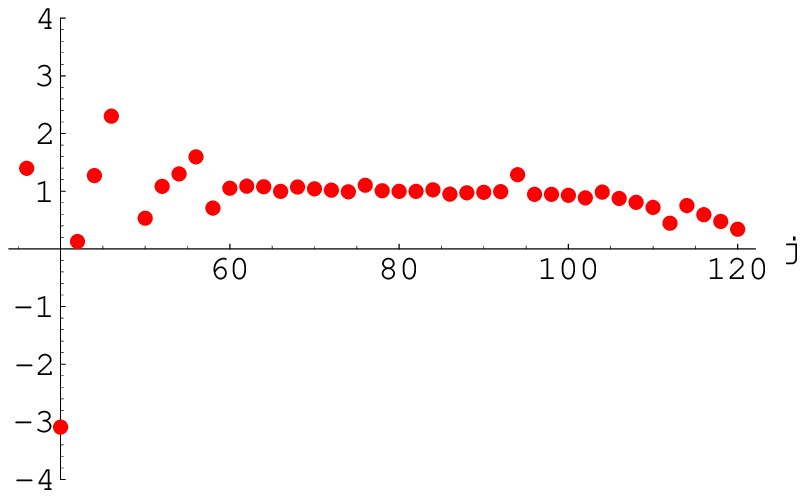} \hspace{1cm}
\includegraphics[width=7cm]{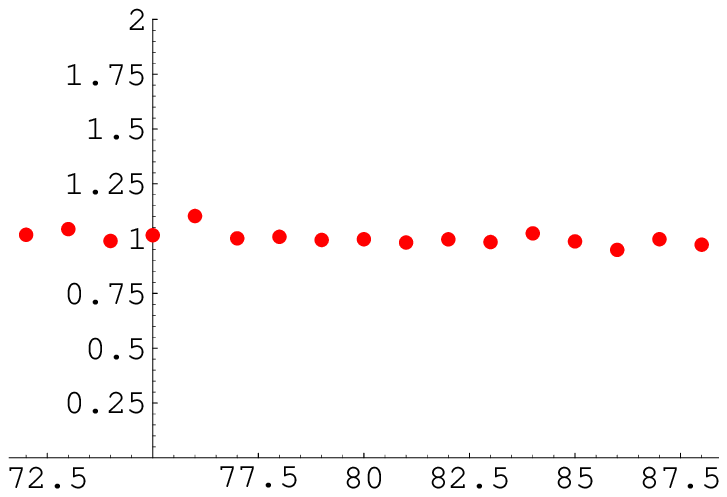} \end{center}
\caption{\footnotesize Numerical study of the ratio between \Ref{4} (with $\phi$ giving the positive 
eigenvalue) and \Ref{ansatz} in the equilateral $j_0=j$ case, with $j=80$. 
On the left panel, a wide range of admissable values of $a$ 
is shown. The behaviour is constantly around 1 only in a region with $a$ around 80. 
For very small $a$, we enter the $V^2\simeq 0$ regime,
and \Ref{4} is not anymore equal to \Ref{ansatz}, as shown by the presence of the oscillations. Similarly
the $a\simeq 2j$ regime shows a (still oscillating) exponential decaying.
The region where the approximation is valid is shown in the right panel. 
\label{giuly}}
\end{figure}

These results can be confirmed numerically, see Fig.\ref{giuly}. We consider the simple equilateral configuration
$j_0=j$, for which notice that $\cos\theta_0 = \cos\theta = - 1/3$.

Before concluding this section, some comments are in order.

\begin{enumerate}
\item
We started with a Gaussian with a generic width $\sigma$. Requiring the stability of this Gaussian under the quantum evolution leads to a \emph{unique choice} for the width. So the width can be fixed dynamically.
This is independent of the phase term in \Ref{ansatz}.

Furtermore, notice that this choice is precisely the one made in \cite{Io,3d} in order to obtain 
the correlation of an harmonic oscillator.
Here we show that this choice arises naturally from a dynamical requirement.

\item
The choice of a real phase (the cosine)
is rather natural given the form of the kernel asymptotics \Ref{asymp}. On the other hand, a complex phase
of the type $\exp\{i \theta d_j/2\}$ has been more commonly used in the literature. This choice has the
advantage of selecting only one of the two exponentials in \Ref{asymp}, thus introducing a notion of
``observer-induced'' orientation.

Concerning this choice, the reader can easily check that an ansatz of the type
\equ\label{ansatz1}
\Psi_{q}(a) = N \, \exp\{-\f1{4\sigma}(a-j_0)^2 + i \f{d_a}2 \theta_0 + i \phi \}
\nequ
can not solve the eigenvalue problem \Ref{eigen}. On the other hand, one can show proceeding as above that
\Ref{ansatz1} with $\sigma$ given in \Ref{sigma} and $\phi$ in \Ref{plus} satisfies the equation
\equ
\sum_a K[a,b,T] \, \Psi_q(a) = \overline{\Psi_q(b)}.
\nequ
We can read the above equation as a stability condition where the state is turned into its complex conjugate.
This can be also compared with the semiclassical tetrahedron used in \cite{semi}.

\item
As shown in \cite{3d}, the modified Bessel functions of the first
kind $I_n(z)$ satisfy
\equ
\f{ I_{|j-j_0|}(2\sigma{j_0})-I_{j+j_0+1}(2\sigma{j_0})}
{\sqrt{ I_{0}(4\sigma{j_0}) - I_{2j_0+1}(4\sigma{j_0})}} \simeq
N \exp\{-\f1{4\sigma}(a-j_0)^2\}.
\nequ
This means that also this Bessel state approximately solves the eigenvalue problem.
However, it does not solve it exactly.
The advantage of this boundary state is that it admits a simple SU(2) Fourier transform,
which was used in \cite{grav3} to cast the 4d graviton into an integral representation.

\end{enumerate}

The consequences of choosing the physical boundary state computed here in the evaluation
of the graviton propagator are discussed below in Section 5.

\section{Physical state in the general boundary formalism}
In this section we discuss a more speculative construction. Above we computed the leading order
approximation of a state that (i) solves \Ref{evo} and (ii) is peaked around a semiclassical
geometry, but we were not able to extend this result to the non-perturbative theory. However, the
situation becomes somewhat simpler in the so-called general boundary formalism, where also the four
bulk edges are varied freely. Then the boundary state has to carry information on the background
value of the (intrinsic and extrinsic) geometry of all six edges, and correlations between all six
of them can be computed. Such a general boundary formalism has been advocated to be the most
natural area for quantum gravity \cite{GB}, and is indeed the setting used for the 4d spinfoam
graviton calculations \cite{RovelliProp,grav2,grav3,4dnum}.

In general we might expect the general boundary problem to be more complicated to deal with than the
fixed-time setting. Yet in the particular 3d case the fact that the theory is topological
strongly simplifies the analysis, because once the topology and the triangulation are fixed,
there is a single physical state.
Assuming trivial topology, the latter is given in the group representation by
\equ\label{generic}
\psi = \prod_{{\rm ind} f} \d(g_f)
\nequ
where $g_f = {\prod}_{l\in \p f} g_l$ represents the gravitational holonomy on a closed path $\p f$.
The product is over the independent faces, 
and the condition $F=0$ is ensured everywhere.
The spin representation of \Ref{generic} can be obtained projecting it on the spin network basis
\equ
\Psi_{\{j_l\}}(g_1, \ldots g_6) = \prod_{n} i_{n} \, \prod_{l} D^{(j_l)}(g_l).
\nequ

\begin{floatingfigure}[r]{5cm}
\includegraphics[width=3cm]{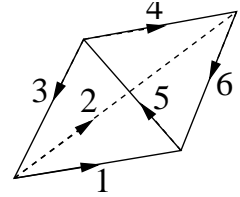}
\caption{\footnotesize Orientation.\label{anelka}}
\end{floatingfigure}
In the case considered here, the triangulation is given by a single tetrahedron, and \Ref{generic}
is a product of four deltas ensuring the flatness of each face.
Only three faces are independent, so we can get rid of one delta.
We fix the orientation of the edges as in the adjacent figure.
Choosing to neglect $\d(g_1 g_5 g_3)$ we can simply write the state
\equ\label{psitet}
\wtl{\psi}_0(g_1, \ldots g_6) = \d(g_3 g_2 g_4^{-1}) \, \d(g_5 g_4 g_6) \, \d(g_2 g_6 g_1^{-1}).
\nequ
This state is physical: it satisfies $F=0$ everywhere and it is thus a solution of the Hamiltonian constraint.

To obtain the spin representation of this state, we project it on the spin network basis
\equ\nonumber
\Psi_{\{j_l\}}(g_1, \ldots g_6) = \sum_{m_i, n_i}
D_{m_1 n_1}^{(j_1)}(g_1) \ldots D_{m_6 n_6}^{(j_6)}(g_6) \, \left(m_1 m_2 n_3 \right)
\, \left(n_1 m_5 n_6 \right)\, \left(m_6 n_4 n_2 \right)\, \left(m_4 n_5 m_3 \right),
\nequ
where the $\left(m_1 m_2 m_3 \right)$ are normalized Clebsch-Gordan coefficients.
A straightforward calculation then gives the projection
\eqa\label{physGB}
\psi_0(j_1 \ldots j_6) = \int dg_1 \ldots dg_6 \ \wtl{\psi}_0(g_1, \ldots g_6) \
\Psi_{\{j_l\}}(g_1, \ldots g_6) = \left\{ \begin{array}{ccc} j_1 & j_2 & j_3 \\ j_4 & j_5 & j_6 \end{array}\right\}.
\neqa
We see that the physical boundary state coincides with the kernel, so that correlations 
(see next Section) now read:
\equ\label{Wsq}
W_{ab} = \f1{j_0^4} \f1{\cal N} \sum_{j_l} {\mathbbm h}(j_a) \, {\mathbbm h}(j_b) \, \prod_j d_j \,
\left\{ \begin{array}{ccc} j_1 & j_2 & j_3 \\ j_4 & j_5 & j_6 \end{array} \right\}^2.
\nequ
This result might look surprising at first, but it can be understood as follows. The boundary state 
$\psi_0(j_1 \ldots j_6)$ is the state induced by the (exterior) bulk geometry onto the tetrahedron. We have assumed
a trivial boundary topology (homomorphic to ${\cal S}^2$) and a trivial bulk topology. Therefore we
obtained a spinfoam amplitude in $\{6j\}^2$ which is naturally associated to the triangulation of
the closed ${\cal S}^3$ manifold with two tetrahedra. In this context it is a natural result.

Notice that to go back to the time-gauge setting used in the previous Section it suffices to
gauge-fix four opposite representations of the spin network state to $j$. We can interpret this
gauge-fixed spin network state with the four representations fixed to $j$ as if it was induced by a
non-trivial bulk topology. Indeed, we can imagine topological defects on each of these edges
inducing non-trivial holonomies around these edges. These are in turn interpreted as particles
traveling along these edges and $j$ is the proper time along the particle trajectory \cite{PR,
altri}. From this perspective, the gauge-fixed spin network state with fixed $j$ appears as a
physical state for a non-trivial topology, or, equivalently, for the coupled system ``gravity +
particles".

\section{Graviton propagator revisited}
The results presented so far give new inputs to the physical interpretation of the 3d graviton
propagator computed in \cite{Io,3d}.
Let us recall that in the time-gauge setting, \Ref{Wq} is realized as \cite{Io}
\equ\label{W3d}
W(j_0, j) = \f1{j_0^4} \f1{\cal N} \sum_{a,b} {\mathbbm h}(j_a) \, {\mathbbm h}(j_b) \, \Psi_q[a,b] \, K_{ab}[j]
\nequ
where ${\mathbbm h}(j_a)=j_a(j_a+1)-j_0(j_0+1)$ represents a field insertion.
This quantity gives the quantum correlator between the two opposite edges, and formally can be considered
as a projected component of the graviton propagator. The kernel is \Ref{Ktet},\footnote{Up
to factors $d_j$ irrelevant for the leading order analysis.} and in \cite{Io}
the generic Gaussian ansatz with a complex phase \Ref{ansatz1} was used.
Here we want to study what happens when we use the physically determined unique boundary state.
The leading order in the large spin limit can be easily computed following the same procedure
of Section 3, and noticing that ${\mathbbm h}(j_a)\simeq 2 j_0 \d j_a$.
The result is a phased Gaussian integral whose evaluation gives
\equ\label{W3}
W(j_0,j) = \f4{j_0^2}
\f{\sum_{\eps=\pm}  \exp\{i\eps[\f\pi4+2d_{j_0}\th -2\phi] \} \, (A_\eps^{-1})_{12} \, / \, \sqrt{\det A_\eps}}
{\sum_{\eps=\pm} \exp\{i\eps[\f\pi4+2d_{j_0}\th -2\phi] \} \, / \, \sqrt{\det A_\eps}}
\nequ
where the quadratic matrix is
\equ
A_\eps = \f1{2\sigma}\left(\begin{array}{cc} 1-i \eps \cot \th & i {\eps}\f1{\sin\th} \\
i {\eps}\f1{\sin\th} & 1-i \eps \cot \th \end{array} \right),
\qquad
\det A_\eps = \f1{2\sigma^2 \sin\th} \, e^{i \eps (\th - \f\pi2)}.
\nequ
Simple algebra shows that $\phi$ exactly cancels the phases in \Ref{W3}
for both choices \Ref{plus} and \Ref{minus}, and finally we get
\equ
W(j_0,j) = \f{4\sigma}{j_0^2} \, \cos\theta.
\nequ

Following \cite{Io}, this can be interpreted as the real part of the correlator of an harmonic
oscillator of frequency $\om = \theta/T$. However, the fact that we are now working with the real
part allows us to have a direct interpretation in terms of the inverse distance traveled along the
bulk edge. Indeed, using the explicit values \Ref{sigma} and \Ref{coseni} for $\sigma$ and
$\cos\theta$ we have
\equ\label{Wfinal}
W(j_0,j) = -\f1{2j},
\nequ
independently of $j_0$ (although the very definition of the leading order approximation is based on the
presence of the background $j_0$). This remarkable result shows explicitly at leading order that the use of a
physical boundary state leads to the correct behaviour of the correlations.

Yet to make a more precise connection between this correlation and the (projections of the free)
graviton propagator, there are two crucial issues that have to be addressed: in the continuum theory
the 3d propagator is (i) gauge-dependent, and (ii) is a \emph{pure gauge} quantity, consistently with
the absence of local degrees of freedom of 3d GR. These issues were not addressed in \cite{Io,3d},
and the question remained open for some time.

The answer to point (i) is rather simple: the expression \Ref{W3d} is defined in the time-gauge
obtained fixing the bulk edges to $j$. Consistently also the physical boundary state is evaluated in the time gauge.
Now in the continuum the time gauge is only a partial gauge fixing (e.g. \cite{Mattei}), and one needs
a further spacelike gauge fixing (such as the Coulomb gauge) in order to compute the
the 2-point correlation, which involves the insertion of the gauge-dependent quantities $h_{ab}(x)$.
In \Ref{W3d} on the other hand, the correlations are studied via the gauge-independent observables
$\mathbbm h$ (they are $\SU(2)$ Casimirs), thus the time gauge is a full gauge-fixing, and the correlation \Ref{Wfinal}
depends only on it.

Answering point (ii) is on the other hand more subtle.
This question was addressed in \cite{bianca}, where it was shown in 3d Regge calculus on a infinite
regular lattice that the propagator is a pure gauge, namely that the propagating quantity in the correlations
comes entirely from the gauge-fixing term in the discrete action. It would be useful to extend those results
to the spinfoam formalism considered here, however that
calculation relies on the fact that one is working on an infinite lattice, and thus gauge degrees
of freedom can be identified as translations on the lattice.
This is not possible from the viewpoint of a single tetrahedron. As shown in \cite{bianca}, the left over of
gauge invariance on a single tetrahedron can be identified at the perturbative level as the global symmetry
corresponding to the following global rescaling on the background $\ell_e$,
\equ\label{rescale}
\d\ell_e \mapsto \d\ell_e + N \ell_e.
\nequ
When we choose the time gauge we clearly break this symmetry. However one has to show that it is
precisely this gauge-fixing term that propagates in \Ref{Wfinal}, and nothing else. To do so, we
consider the general boundary formalism discussed above, where the symmetry \Ref{rescale} is
present.

In the general boundary formalism, using the physical boundary state \Ref{physGB} leads to the formula
\Ref{Wsq} for the correlations between edges. The perturbative expansion of this quantity can be studied again using
\Ref{asymp}, which gives in the large spin limit
\equ\label{asy2}
\{6j\}^2 \sim \f1{24\pi V(j_e)} \, \Big[1 - \sin\Big(2 S_{\rm R}(j_l)\Big) \Big].
\nequ
The constant first term in the square bracket shows immediately that the summand is divergent,
consistently with the fact that we have not fixed the gauge. Furthermore, notice that the dynamics
is only encoded in the oscillatory second term. Imagine now to gauge-fix the symmetry
\Ref{rescale} introducing \`a la Faddeev-Popov terms like $\exp\{-\d j^2\}$. If this gauge-fixing term has a non-diagonal structure
then it will give the dominant contribution to the correlations, with the Regge action in
\Ref{asy2} exponentially suppressed. In this case, we can clearly claim that at least
perturbatively the correlations are pure gauge. If on the other hand the gauge-fixing term has no
off-diagonal structure, then the correlations would indeed come from the dynamical Regge term, but
because of its oscillatory nature they would be completely suppressed at large scales. They would then 
be some sort of quantum noise of no semiclassical counterpart.

\section{Conclusions}

We considered the stability condition for physical states in quantum gravity. Within the simple toy
model of a 3d Riemannian tetrahedron, we realized explicitly this condition as an eigenvalue
problem in the spinfoam formalism, given in \Ref{eigen}. 
We wrote this problem as an integral equation (see Appendix A), and found some
exact solutions. At leading order in the large spin limit we were able to find a class of solutions
which can also be interpreted as semiclassical states for the linearized theory around a fixed
background. This class of solutions are Gaussian wavefunctions \Ref{ansatz} with a width that is dynamically
fixed to the value given in \Ref{sigma} by the stability condition. 
This result positively answers the question that one can fix the
width of this semiclassical states looking at the spinfoam dynamics. Furthermore, in this specific
toy model, the amplitude fixed in this way turns out to be exactly the one giving correlations
which are at leading order in agreement with the discretization of the continuum linearized quantum
theory. Finally, our results show that the PR model supports the dynamical propagation of stable
wave packets of geometry.

\section*{Acknowledgements}
Research at Perimeter Institute for Theoretical Physics is supported in
part by the Government of Canada through NSERC and by the Province of
Ontario through MRI.

\appendix

\section{Analyzing the $\{6j\}$ kernel}
We start from the integral expression \Ref{integral}.
Expressing the group elements $g_{1,2}$ in term of their class angles $\phi_{1,2}$ and their
rotation axis $\hat{u}_{1,2}\in {\cal S}^2$, this becomes:
\begin{equation}\label{Cicciocev}
\left\{
\begin{array}{ccc} a & j & j \\ b & j & j \end{array} \right\} = \f2{\pi^2}
\int_0^\pi d\phi_1 d\phi_2\sin^2\phi_1\sin^2\phi_2\,
\chi_{a}(\phi_1) \, \chi_{b}(\phi_2) \, \int_{-1}^{1} dy\, \chi_{j}(\phi_-) \, \chi_{j}(\phi_+),
\end{equation}
with the mixed class angles defined in term of $y\equiv \hat{u}_1\cdot\hat{u}_2$ as:
\equ
\cos\phi_\pm \,=\,
\cos\phi_1\cos\phi_2 \mp y \sin\phi_1\sin\phi_2.
\nequ
With $y=\cos\alpha$, notice this can re-written as:
\eqa
\cos\phi_- &=& \cos^2\f\alpha2\cos(\phi_1-\phi_2)+\sin^2\f\alpha2\cos(\phi_1+\phi_2),
\no\nonumber
\cos\phi_+ &=& \sin^2\f\alpha2\cos(\phi_1-\phi_2)+\cos^2\f\alpha2\cos(\phi_1+\phi_2),
\neqa
which shows that the angles $\phi_\pm$ run between $(\phi_1-\phi_2)$ and
$(\phi_1+\phi_2)$. Notice also the constraint between the angles:
\equ
\cos\phi_- + \cos\phi_+ \,=\, \cos(\phi_1-\phi_2) + \cos(\phi_1+\phi_2)
\,=\,2\cos\phi_1\cos\phi_2.
\nequ
Using the relation between the $\SU(2)$ characters and the Chebyshev polynomials of the second
type, $\chi_j(\phi)=\sin (2j+1)\phi/\sin\phi=U_{2j}(\cos\phi)$, we can re-express the
$\{6j\}$ as an integral of products of Chebyshev polynomials over $x_{1,2}=\cos\phi_{1,2}$ and
$y$. For instance, we have:
\equ\nonumber
\left\{
\begin{array}{ccc} a & j & j \\ b & j & j \end{array} \right\} = \f2{\pi^2}
\int_{-1}^1 dx_1dx_2dx_-dx_+\,
\delta(x_-+x_+-2x_1x_2)\,
U_{2a}(x_1)U_{2b}(x_2)U_{2j}(x_-)U_{2j}(x_+).
\nequ
%
Expression \Ref{Cicciocev} suggestes to study the eigenvalue problem \Ref{eigen} in terms of the
following Fourier transform of $\psi_a$,
\equ
\what{\psi}(\phi)=\sum_{a\in\N}\sqrt{d_a} \, \chi_a(\phi)\,\psi_a.
\nequ
The eigenvalue equation then becomes a convolution equation,
\equ
\what{\psi}(\phi)=\pm\int_0^\pi d\phi_2 \, \sin^2\phi_2 \, {\widetilde K}(\phi,\phi_2,j) \, \what{\psi}(\phi_2),
\nequ
with kernel given by
\equ
\widetilde{K}(\phi,\phi_2,j) = \int_0^\pi d\phi_1 \, \sin^2\phi_1 \,
K(\phi_1,\phi_2) \, \wtl{\delta}(\phi,\phi_1),
\nequ
where
\equ
K(\phi_1,\phi_2)\,=\,\int_{-1}^1 dy\,\chi_{j}(\phi_-) \, \chi_{j}(\phi_+),\qquad
\wtl{\delta}(\phi,\phi_1)\,=\, \sum_a d_a \chi_a(\phi_1)\chi_a(\phi).
\nequ
$K(\phi_1,\phi_2)$ can be evaluated exactly writing it in term of Chebyshev polynomials
while $\wtl{\delta}(\phi,\phi_1)$ is a distribution over $\SU(2)$ which can be directly
evaluated using the explicit expression for the characters,
$\chi_a(\phi)=\sin(2a+1)\phi/\sin\phi$.

Exact solutions to the convolution equation are still under investigation. At least, we already
know that gaussian states on $\SU(2)$ will approximatively solve it in the asymptotic limit for
large spins since they are the Fourier transforms of the gaussian states on the representation
labels.


\section{Extremal cases}
The first and last rows of the matrix $K_{ab}[j]$ admit simple expressions, that we report here
for completeness and future use. Notice that these are cases for which $V^2$ is not positive so the asymptotics
\Ref{asymp} do not hold.

\bigskip

\underline{\emph{Case $b=0$}}

\begin{figure}[ht]\begin{center}
\includegraphics[width=5cm]{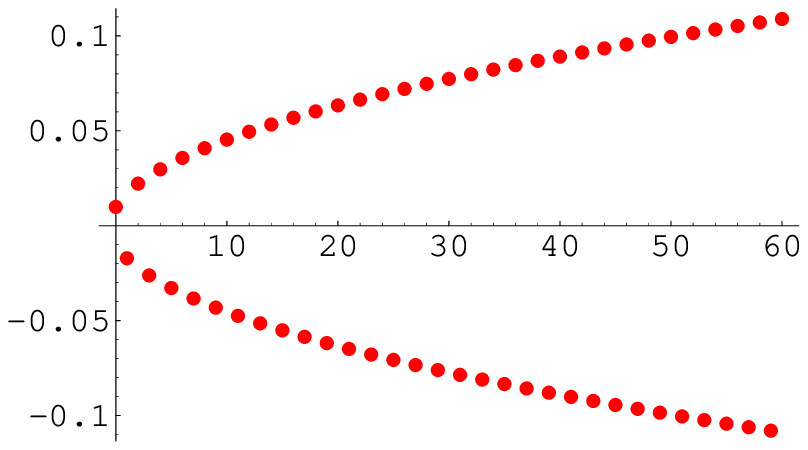} \hspace{1.5cm}
\includegraphics[width=5cm]{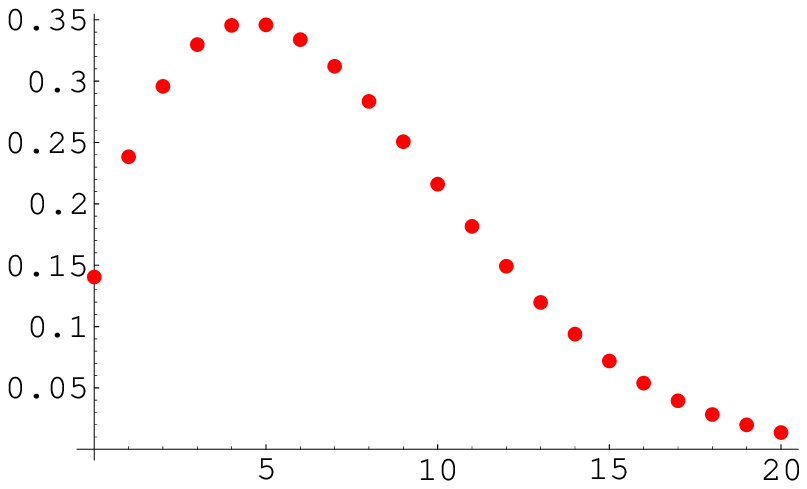}\end{center}
\caption{\small{\emph{Left panel}. A typical plot for $g_j(a)$. Here $j=50$, and for clarity only
part of the full range $a\in[0,100]$ is shown.
\emph{Right panel}. A typical plot for $f_j(a)$. Here again
$j=50$ and $a\in[0,100]$. The peak is at $\sqrt{25}=5$.}}\label{fig}
\end{figure}

In this case we have $V^2=0$ and
\equ
\left\{\begin{array}{ccc} a & j & j \\ 0 & j & j \end{array} \right\} \equiv
\f{(-1)^{2j+a}}{d_j}.
\nequ
Therefore $g_j(a) \equiv K_{a0}[j] = (-1)^{2j+a}\f{\sqrt{d_a}}{d_j}$.
This is a rather simple expression, plotted on the left panel of Fig.\ref{fig}.

For generic $V^2 \simeq 0$ configurations, the asymptotics of the $\{6j\}$ is \cite{Ponzano}
\equ\label{asymp=}
\{6j\} \sim \f1{\sqrt[6]{64 \, S}}\,\Big[\cos\Phi \,{\rm Ai}(z)+\sin\Phi \,{\rm Bi}(z)\Big],
\nequ
where $S$ is the four times the product of the four triangle areas, Ai and Bi are Airy functions \cite{Abr}, and
\equ
\Phi = \sum_e (j_e-\f12)\,{\rm Re} \, \th_e, \qquad
z = \left\{ \begin{array}{lc} -(3 V)^2\,S^{-\f23}, & V^2>0 \\
(3 |V|)^2\,S^{-\f23}, & V^2<0 \end{array} \right.
\nequ

\bigskip

\underline{\emph{Case $b=2j$}}

In this case we have $V^2= -\f{j^2a^4}{36}<0$ and
\equ
\left\{\begin{array}{ccc} a & j & j \\ 2j & j & j \end{array} \right\} \equiv
\f{4j^2 \, \Gamma(2j)^2}{p(a,j) \, \Gamma(2j+a) \, \Gamma(2j-a)}
\nequ
where
\equ
p(a,j) = (2j+a)(2j-a)(2j+a+1).
\nequ
The function $f_j(a) \equiv K_{a 2j}[j]$ has a more interesting behaviour, and a peak
approximately at $\sqrt{j/2}$, see the right panel of Fig.\ref{fig}.
The position of the peak can be computed analytically:
\equ\label{der}
\p_a W_j(a) = -4j^2 \, \sqrt{4j+1}\, \Gamma(2j)^2 \,
\f{q(a,j)-(2a+1) \, p(a,j) \,[\psi_0(2j-a)-\psi_0(2j+a)]}{\sqrt{2a+1}\, p(a,j)}
\nequ
where
\equ
q(a,j) = 5a^3 + 8j^3+6a^2(j+1)+2a-4aj(j-1)
\nequ
and
\equ
\psi_0(n)=-\gamma_{\rm E}+\sum_{k=1}^n\f1k
\nequ
is the digamma function. Its power series for large values of the argument starts with
$\ln(n-1)+1/2(n-1)$. Using this approximation, the leading order for big $j$ of the numerator of
\Ref{der} vanishes at
\equ
2j-a(4a+3)=0
\nequ
from which
\equ
a = \f18\Big(\sqrt{9+32 j}-3\Big) \sim \sqrt{\f{j}{2}}.
\nequ

For generic $V^2 < 0$ configurations, the asymptotics of the $\{6j\}$ is \cite{Ponzano}
\equ\label{asymp<}
\{6j\} \sim \f1{2\sqrt{12\pi V}}\,\cos\Phi \,e^{-|\Omega|},
\nequ
with
\equ
\Phi = \sum_e (j_e-\f12)\,{\rm Re} \, \th_e, \qquad
\Omega = \sum_e j_e \,{\rm Im} \, \th_e.
\nequ



\end{document}